\documentclass[twocolumn,aps,prb,floatfix,showpacs]{revtex4}
\usepackage{graphicx}
\usepackage{dcolumn}
\usepackage{bm}
\usepackage{psfrag}
\usepackage{amssymb,amsmath}

\begin{document}
  \title{Bound states and persistent currents in topological insulator rings}
  \author{Paolo~Michetti and Patrik~Recher}
  \affiliation{Institute of Theoretical Physics and Astrophysics, University of W\"urzburg, D-97074 W\"urzburg, Germany}
  \date{\today}

 \begin{abstract}
   We analyze theoretically the bound state spectrum of an Aharonov Bohm (AB) ring in a
   two-dimensional topological insulator using the four-band model of HgTe-quantum wells as a concrete example. 
   We calculate analytically the circular helical edge states and their spectrum as well as the bound states evolving 
   out of the bulk spectrum as a function of the applied magnetic flux and dimension of the ring.
   We also analyze the spin-dependent persistent currents, which can be
   used to measure the spin of single electrons. 
   We further take into account the Rashba spin-orbit interaction which mixes the spin states and derive its effect on 
   the ring spectrum.
   The flux tunability of the ring states allows for coherent mixing of the edge- and the spin 
   degrees of freedom of bound electrons which could be exploited for quantum information processing 
   in topological insulator rings.
 \end{abstract}

  \maketitle

  \section{Introduction}
  Mesoscopic rings offer an ideal testbed for the investigation of geometrical phase effects 
  in quantum mechanics like the well known Aharonov-Bohm (AB) effect~\cite{aharonov1959}.
  Experimental observation of the energy spectra of quantum rings, showing clear signatures of AB periodicity, 
  have been obtained by means of optical spectroscopy~\cite{lorke2000, bayer2003}. 
  The predicted persistent currents~\cite{buttiker1983} have been experimentally measured, 
  also on a single ring, in various experiments in metals~\cite{chandrasekhar1991,bluhm2009,bleszynski2009} 
  and semiconductors~\cite{mailly1993}.
  Phase effects, such as the Aharonov-Casher effect~\cite{konig2006,borunda2008} and the 
  Sagnac phase shift~\cite{zivkovic2008} have been studied.
  Although many of these features can be understood in terms of a single particle picture, 
  quantum rings are also attractive laboratories, being quasi-1D systems, 
  to study many-body effects. A review about this topic can be found in Ref.~\onlinecite{viefers2004}.
  In conjunction with spin-orbit interaction, quantum rings 
  are an active field of research for 
  spin manipulation~\cite{berche2010}. 
  Proposed devices, based on a ring geometry, 
  include spin-filters~\cite{popp2003}, spin beam splitters~\cite{foldi2006}, spin current parametric pumping~\cite{citro2006}, 
  spin qubits~\cite{foldi2005}, or quantum rings which can be used to compose networks for spintronic 
  applications~\cite{foldi2009}.

  Topological insulators (TI) are a new class of time-reversal invariant materials that 
  have a bulk gap and protected surface states. 
  These materials are distinguished from a trivial insulator (with no topological protected 
  surface states) by a $Z_2$ topological number~\cite{kane2005b}.
  In two dimensions, these materials exhibit the quantum spin Hall effect~\cite{kane2005,bernevig2006} 
  and typically are small gap semiconductors, with the peculiarity that their  
  gap, originated from strong spin-orbit coupling, is inverted with respect to that of a 
  normal insulator, allowing for the presence of  
  topologically protected edge states~\cite{kane2005,bernevig2006,xu2006}.
  At each boundary two gapless counterpropagating edge modes exist that have opposite 
  spin and are related by Kramer's theorem, known as helical edge modes.
  A successful realization of a  two dimensional TI phase has been obtained in 
  HgTe/CdTe quantum wells, where the topological phase has been confirmed with the observation 
  of the ballistic helical edge channels~\cite{konig2007,konig2008}.
  TIs, with their dissipationless edge states, represent a new promising route to 
  spin manipulation in semiconductors.

  A ring geometry in a TI with its well localized wavefunctions is optimal for spin manipulation, 
  using a magnetic flux.
  It has been proposed, on the basis of numerical transport calculations, that the AB effect in a 
  topological insulator disc can be exploited to perform spin rotation and spin filtering~\cite{jiang2009},  
  or to create a topological spin transistor~\cite{maciejko2010}.   
  Similar properties are expected in a general closed disk geometry connecting 
  two quantum point contacts in which coherent oscillations in the magnetoconductance are expected 
  due to the AB effect~\cite{chu2009}.
  It is clear that, if experimentally confirmed, these properties can make the disk, ring and hole geometry 
  extremely interesting also from an application point of view.

\begin{figure}[bt]
    \centering	
    \includegraphics[width=5.cm]{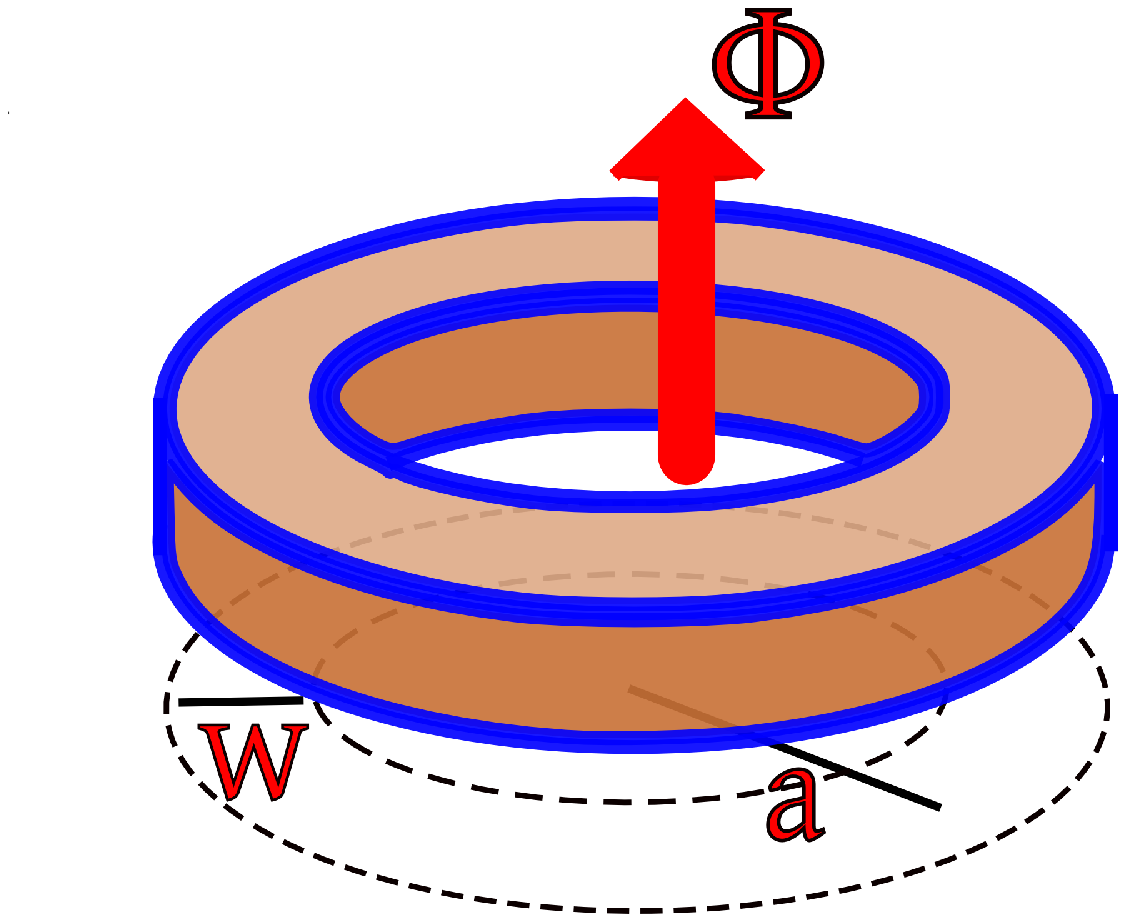}
    \caption{(Color online) Scheme of a topological insulator ring of radius $a$ and width $w$.
      A magnetic field, generating a flux $\Phi$ is threading the ring.
    }
    \label{fig:ring}
  \end{figure}

  In this paper we consider the boundstates of a ring made from a two-dimensional TI (see Fig.~\ref{fig:ring}). 
  The boundstates are found analytically as a function of a magnetic flux, piercing the ring hole, 
  and dimensions of the ring in the topological non-trivial (edges states and bulk states) 
  and topological trivial (only bulk states) regime. 
  We show that the specific bandstructure of such TIs allows for manipulations of 
  electron spin states using the magnetic flux and their read-out through a spin-selective 
  persistent current in the ring.

  In addition, the special topology of a ring introduces a 
  new type of two-level system consisting of inner and outer-edge localized bound states where 
  their mixing can be tuned by the flux. 
  Together with the spin of an electron, these features allow for new ways to coherently manipulate 
  quantum states in topological insulators.

  The paper is organized as follows.
  In section II we solve the edge and bulk states of a TI in cylindrical coordinates, 
  described by an effective four-band model for HgTe/CdTe quantum
  wells~\cite{bernevig2006}, exactly.
  We summarize here, for the benefit of those who are not intersted in the more mathematical details, 
  the main results of section II.

  The total out-of-plane angular momentum is a conserved quantity of the system and we associate the half-odd integer quantum number $m$ to it, this fact permits us to separate azimuthal and radial variables in the spinor [Eq.~(\ref{eq:spinor})].
  We can express the radial part of the spinor components [Eq.~(\ref{eq:comp1}) and (\ref{eq:comp2})] in general  
  as a combination of ordinary and modified Bessel functions, both for the bulk [Eq.~(\ref{eq:bulk})] and 
  edge modes [Eq.~(\ref{eq:edge})].

  We find the eigenstate spectrum by assuming vanishing boundary conditions for a ring, a disk or a hole geometry, 
  obtainining transcendental secular equations, that we numerically solve for the 
  eigenenergies, and eventually we obtain the corresponding eigenfunctions.
  So in general an eigenstate of the system can be labelled as $|n,m,\tau\rangle$, 
  with $n$ the radial quantum number, associated with the number of nodes of the radial wavefunction, 
  the total angular momentum $m$ and the spin $\tau=\pm1$.

  In Section III we show the results of our calculation, in particular focusing on the 
  bound states of a ring subject to the AB effect.
  We present the edge dispersion for a disk, a hole and a ring, where a minigap arises due to the confinement, 
  which exponentially shrinks with increasing ring thicknesses.
  We show that the sign of the Dirac mass, which distinguishes the topologically 
  trivial from the topologically non-trivial insulator, is important to the eigenstates in the gap 
  but also in the bulk spectral region.
  We also calculate the persistent currents as a function of the magnetic flux.

  In Section IV we consider the inclusion of Rashba spin-orbit interaction which couples the two spin-blocks. 
  The main effect, for a sufficiently small interaction, is the opening of a spin splitting gap 
  at specific values of the magnetic flux.
  As a consequence, we can perform coherent spin manipulations with the magnetic flux.

  In Section V we present a scheme for the coherent manipulation of quantum states stored locally in 
  the spin or the edge degrees of freedom of a TI ring, by means of an AB flux.

  \section{Four band model of Hg-Te quantum wells}
  For a realistic description of a two-dimensional topological insulator, we 
  use the effective four-band model~\cite{bernevig2006,konig2008} 
  for HgTe quantum wells, composed out of the $E_1$ and $H_1$ subbands, 
  and the electron spin degree of freedom $|\pm\rangle$ (see Appendix~\ref{app:spin} for a discussion 
  of these basis states and the meaning of spin). 
  The four-band Hamiltonian is built with respect to the basis $\big\{|E_1+\rangle$, 
  $|H_1+\rangle$, $|E_1-\rangle$, $|H_1-\rangle\big\}$ (see Eq.~(\ref{eq:gamma})), 
  where the $E_1$ states are a mixing of the s-like $\Gamma_6$ band with the $\Gamma_8$ light-hole 
  band, and $H_1$ represents basically the $\Gamma_8$ heavy-hole band. 
  In this basis, $\pm$ stands for the sign of the total angular momentum projection of the $E_1$ 
  and $H_1$ bands, i.e. the sum of spin and orbital angular momentum.
  In the absence of inversion symmetry breaking terms (as the Rashba interaction discussed in Section IV), the Hamiltonian is separable into two blocks related by time-reversal symmetry~\cite{bernevig2006}
  \begin{equation}
    \label{H1}
    {\tilde H}=\left(\begin{array}{cc}h(k) &  0 \\ 0 & h^{*}(-k)\end{array}\right)
  \end{equation}
  with $h(k)=\epsilon(k){\rm I}_{2\times2}+d_{a}(k)\sigma^{a}$ and
  $d_{a}(k)=(Ak_{x},-Ak_{y},M(k))$ where $\sigma^{a}$ is the vector
  of Pauli matrices. In Eq.~(\ref{H1}), $\epsilon(k)=C-Dk^2$, and the Dirac mass
  term $M(k)=M-Bk^2$ with $k^2=k_x^2+k_y^2$.

  The parameters $A,B,C,D,M$ depend on the QW geometry. 
  Realistic estimates can be made by a comparison of the effective model with a well 
  established $8 \times 8$ Kane Hamiltonian~\cite{novik2005}. 
  In the following we will use the parameter values $A=375$ meV nm, $B=-1.120$ eV ${\rm nm}^2$ 
  and $D=-730$ meV ${\rm nm}^2$. 
  The parameter $C$ results only in a shift of the energy and we can set $C=0$ without loss of generality. 
  The Dirac rest mass $M$ depends on the QW thickness and $M<0$ corresponds to the inverted (topologically 
  non-trivial) regime whereas  $M>0$ corresponds to the normal (topologically trivial) regime.

  This model is able to describe both the edge states and the bulk states in a reasonable energy 
  range \cite{schmidt2009}.
  We note that Eq.~(\ref{H1}) is not restricted to HgTe QWs but describes also other two-dimensional 
  TIs like thin films of ${\rm Bi_{2}Te_{3}}$ and ${\rm Bi_{2}Se_{3}}$.~\cite{Liu2009}
  For the ring, it is useful to express Eq.~(\ref{H1}) in polar coordinates $x=r \cos\theta$, $y=r \sin\theta$.

  We now use $\vec{k}=-i\vec{\nabla}$ and perform the unitary transformation 
  \begin{equation}
    U=\frac{1}{2}\left[ (\tau_0-\tau_z)\sigma_y + (\tau_0+\tau_z)\sigma_0\right],
    \label{eq:trans}
  \end{equation}
  which changes the old basis into $\big\{|E_1+\rangle$, $|H_1+\rangle$, $-i|H_1-\rangle$, $i|E_1-\rangle\big\}$,
  we obtain the transformed Hamiltonian,
  \begin{equation}
    H =  C + M \tau_z\sigma_z + (D\mathbb{I} + B \tau_z\sigma_z)\mathbf{\Delta}
    -i A  e^{i \sigma_z \theta} \mathbf{\Pi}
    \label{eq:ham_sym_tot}
  \end{equation} 
  with  
  \begin{eqnarray}
    \mathbf{\Delta}(r,\theta)&=&\left(
      \frac{\partial^2 }{\partial r^2}
      +\frac{1}{r}\frac{\partial }{\partial r}
      +\frac{1}{r^2}\frac{\partial^2 }{\partial \theta^2}
    \right)\\
    \mathbf{\Pi}(r,\theta)&=&\left(
      \frac{\partial }{\partial r} \sigma_x
      - \frac{1}{r} \frac{\partial }{\partial \theta} \sigma_y 
    \right),
  \end{eqnarray}
  where $\tau_a$, $\sigma_a$ represent the Pauli matrices for the spin (Kramer's partners $\pm$) 
  and pseudo-spin ($E1$, $H1$) components, respectively.
  %
  %
  %
  %

  In the following table we define the orbital angular momentum $\hat{l}_z$, the \emph{pseudo} spin 
  operator $\hat{S}_z$ and the time reversal operator $\hat{T}$ for the unsymmetrized and the symmetrized basis.
  \begin{equation}
    \begin{tabular}{|c|c|}
      \hline
      ${\tilde H}$ &  $ H$   \\
      \hline
      $\hat{l}_z = -i \hbar \partial_\theta$ & $\hat{l}_z = -i \hbar \partial_\theta$ \\
      $\hat{S}_z = \frac{\hbar}{2} \tau_z \sigma_z$  & $\hat{S}_z = \frac{\hbar}{2} \tau_0 \sigma_z$ \\
      $\hat{T} = i \tau_y \sigma_0  \hat{K}$      & $\hat{T} = -  \tau_x \sigma_y \hat{K}$\\
      \hline
    \end{tabular}  \label{tab:1}
  \end{equation}

  We further introduce a total angular momentum operator as $\hat{j}_z=\hat{l}_z-\hat{S}_z$ 
  which commutes with $H$  \begin{equation}
    [H,\hat{j}_z] =0.
  \end{equation} 
  The eigenstates of  $\hat{j}_z$ therefore
  can be used for the ring energy eigenstates.
  Consequently,  we can write the spinors for the two blocks ($\tau=\pm1$) of $H$ as 
  \begin{equation}
    \Psi_{m,\tau}(r,\theta) = \frac{e^{i m \theta}}{\sqrt{2 \pi}} 
    \left(\begin{array}{l}
        \chi_1^{m,\tau}(r) e^{i \frac{\theta}{2}}\\
        \chi_2^{m,\tau}(r) e^{-i \frac{\theta}{2}}
      \end{array}
    \right),
    \label{eq:spinor}
  \end{equation}
  where it is easy to verify that 
  \begin{equation}
    \hat{j}_z \Psi_{m,\tau} = \hbar m \Psi_{m,\tau}.
    \label{eq:tot_ang}
  \end{equation}

  Now we need to find the radial wave functions $\chi_{1,2}^{\tau}(r)$.
  We note that the diagonal part of $H$ (putting $\mathbf{\Pi}=0$) is solved by Bessel functions.
  Similarly, if we put $\mathbf{\Delta}=0$, the radial solution for each of the two blocks of $H$ can be expressed 
  in terms of Bessel functions~\cite{recher2007}.
  We consequently use the ansatz $\chi_{1}^{m,\tau}(r) = f_{m+\frac{1}{2}}(K r)$, with $f$ proportional to a 
  function of the Bessel family, defined below 
  (this will result in $\chi_{2}^{m,\tau}(r)\propto  f_{m-\frac{1}{2}}(K r)$ as we will show below).
  To include a magnetic flux $\Phi$ threading the ring hole, we consider the 
  vector potential $\vec{A}=\left(\Phi/2\pi r\right) \hat{e}_{\theta}$.
  Using the minimal coupling procedure $-i\hbar\vec{\nabla}\rightarrow -i\hbar\vec{\nabla}+e\vec{A}$, 
  with $-e$ the electron charge, the inclusion of the magnetic flux is equivalent to 
  \begin{equation}
    \partial_\theta \rightarrow \partial_\theta +i \frac{\Phi}{\Phi_0}
  \end{equation}
  with the magnetic flux quantum $\Phi_0=h/e$.
  The radial wave functions satisfy the Bessel differential equation 
  \begin{equation}
    \mathbf{\Delta}_{\bar{m}\pm\frac{1}{2}} f_{\bar{m} \pm \frac{1}{2}}(K r)= - K^2 f_{\bar{m} \pm \frac{1}{2}}(K r),
    \label{eq:bessel}
  \end{equation}
  with
  \begin{equation}
    \mathbf{\Delta}_{\bar{m}\pm\frac{1}{2}}=
    \frac{\partial^2 }{\partial r^2}
    +\frac{1}{r} \frac{\partial }{\partial r}
    +\frac{1}{r^2} \left( \bar{m} \pm \frac{1}{2}\right)^2,
  \end{equation}
  where $\bar{m}=m+\Phi/\Phi_0$.
  Using $H\Psi_{m,\tau}=E \Psi_{m,\tau}$, we arrive at the following equations
  \begin{equation}
    i \frac{R}{K}
    \left[
      \partial_r
      +\frac{\bar{m}+\frac{1}{2}}{r} 
    \right] \chi_1^{m,\tau} =\chi_2^{m,\tau}
    \label{eq:first}
    \end{equation}
    \begin{equation}
    \left[ 
      \frac{(E\hspace{-0.1cm}+\hspace{-0.1cm}D\hspace{-0.03cm} K^2)^2\hspace{-0.1cm}-\hspace{-0.1cm}(B\hspace{-0.03cm} K^2\hspace{-0.1cm} -\hspace{-0.1cm}M)^2}{A^2}\hspace{-0.05cm} +\hspace{-0.05cm} \mathbf{\Delta}_{\bar{m}+\frac{1}{2}}
    \right]\hspace{-0.05cm} \chi_{1}^{m,\tau}\hspace{-0.03cm}\hspace{-0.1cm} =0,
    \label{eq:second}
  \end{equation}
  with
  \begin{equation}
    R = \frac{A K}{(B K^2 -M)\tau - (E+D K^2)}.
    \label{eq:erre}
  \end{equation}
  Eq.~(\ref{eq:second}) is satisfied, employing the assumption of $\chi_{1}^{m,\tau}$ being a 
  Bessel function (see Eq.~(\ref{eq:bessel})), if
  \begin{eqnarray}
    A^2 K^2=  (E+D K^2)^2 - (B K^2 -M)^2.
  \end{eqnarray}
  Therefore, for a given energy $E$, there are {\it four} possible values for the wave vector $K$
  \begin{equation}
    K^2 = -F \pm \sqrt{F^2 - Q^2},
    \label{eq:modi}
  \end{equation}
  with
  \begin{eqnarray}
    F &=& \frac{A^2 -2(BM+DE)}{2(B^2-D^2)},\\
    Q^2 &=& \frac{M^2-E^2}{B^2-D^2}.
  \end{eqnarray}
  
  Solutions with $K^2>0$ correspond to propagating modes, 
  and $f(r)$ is therefore described by the Hankel functions $H^{(1)}_\nu$ and $H^{(2)}_\nu$.
  For $K^2<0$, instead, $f$ can be expressed with the modified Bessel functions $K_\nu(|K|r)$ and $I_\nu(|K|r)$ 
  describing an evanescent behavior (exponentially growing and decaying) along the radial direction.

  \subsection{Condition for bulk-propagating modes}
  In the spectral range inside the bulk bands ($|E|>|M|$), we expect Eq.~(\ref{eq:modi}) to produce 
  two real ($K= \pm K_1$) and two imaginary $K= \pm i K_2$ solutions.
  Note that for $F^2-Q^2<0$, also complex solutions for $K^2$ are possible. 
  However,  $F^2-Q^2>0$ as long as $M>\frac{A^2}{B}$ which is the case for HgTe QWs.
  Here $K_1$($K_2$) are the modulus of the real(imaginary) solutions of $K$ obtained 
  with the $+$($-$) sign from Eq.~(\ref{eq:modi}). 
  The required condition is that
  \begin{equation}
    F<\sqrt{F^2-Q^2}
  \end{equation} 
  which is satisfied if
  \begin{equation}
    Q^2 <0.
  \end{equation} 
  In HgTe QWs,  $|B|>|D|$, and the condition is automatically satisfied if $|E|>|M|$, as expected.

  Within this spectral region, the first component of the spinor in Eq.~(\ref{eq:spinor}) can be expressed as
  \begin{equation}
    \chi_1^{m,\tau}(r)= \vec{h} \vec{F}_{\bar{m}+\frac{1}{2}}(r),
    \label{eq:comp1}
  \end{equation}
  with 
  \begin{eqnarray}
    \vec{h} &=& \left(h_1, h_2, h_3, h_4\right) \\
    \vec{F}_{\nu}(r) &=& \left( H_{\nu}^{(1)}(\xi_{1}), H_{\nu}^{(2)}(\xi_{1}), I_{\nu}(\xi_{2}), K_{\nu}(\xi_{2})\right),
    \label{eq:bulk}
  \end{eqnarray}
  where $\xi_{i}=K_{i}r$, $i=1,2$.
  The expression for the second component can be found from Eq.~(\ref{eq:first}) 
  and using the following property of the Bessel functions
  \begin{equation}
    \partial_r g_{\nu}(\xi) \pm \frac{\nu }{r} g_{\nu}(\xi) = \pm \eta K g_{\nu-1} (\xi),
    \label{eq:prop_bessel}
  \end{equation}
  with $\eta=+1$ for $H_\nu^{(1)}$, $H_\nu^{(2)}$ and $I_\nu$, and $\eta=-1$ for $K_\nu$.
  We obtain 
  \begin{equation}
      \chi_2^{m,\tau}(r)= i \vec{h} \mathbb{R} \vec{F}_{\bar{m}-\frac{1}{2}}(r),
    \label{eq:comp2}
  \end{equation}
  with the diagonal matrix $\mathbb{R}$ given by 
  \begin{equation}
    \mathbb{R} = {\rm diag}\left(R_{1,+}^{(\tau)}, R_{1,+}^{(\tau)}, R_{2,-}^{(\tau)}, -R_{2,-}^{(\tau)} \right),
  \end{equation}
  where, according to Eq.~(\ref{eq:erre}), we define  
  \begin{equation}
    R_{i,\pm}^{(\tau)} = \frac{A K_i}{(\pm B K_i^2 -M)\tau - (E \pm D K_i^2)},
  \end{equation}
  with $\pm$ depending on whether $K_i$ is the modulus of a real or an imaginary solution of Eq.~(\ref{eq:modi}).

  \subsection{Condition for edge modes}
  In order to have four imaginary solutions of Eq.~(\ref{eq:modi}), in the region $|E|<|M|$, we need, 
  being $|B|>|D|$ and $Q^2>0$, to satisfy the conditions $F>0$ and $F^2-Q^2>0$.
  The latter turns out to be the more restrictive one.
  In particular, if we want to observe 
  the Dirac point of the edge states (of typical value $E=-MD/B$), the material parameters 
  have to satify the following inequality~\cite{zhou2008}
  \begin{equation}
    \frac{A^2}{B^2-D^2}>\frac{4M}{B}.
  \end{equation}
  Again, $K_1$ and $K_2$ are the wave vector moduli of the four modes with imaginary wave vectors  
  obtained from Eq.~(\ref{eq:modi}) with the $+$ and $-$ sign, respectively.

  The spinors for edge states are also obtained from Eq.~(\ref{eq:spinor}) 
  with components described by Eqs.~(\ref{eq:comp1}) and (\ref{eq:comp2}), but with an alternative definition 
  of the basis vector $\vec{F}_{\nu}$ and the diagonal matrix $\mathbb{R}$.
 They are given by
  \begin{eqnarray}
    \vec{F}_{\nu}(r) &=& \left( I_{\nu}(\xi_{1}), K_{\nu}(\xi_{1}), I_{\nu}(\xi_{2}), K_{\nu}(\xi_{2})\right)\label{eq:edge}\\ 
    \mathbb{R} &=& {\rm diag}\left(R_{1,-}^{(\tau)}, -R_{1,-}^{(\tau)}, R_{2,-}^{(\tau)},-R_{2,-}^{(\tau)} \right).
  \end{eqnarray}

  \begin{figure}[tb]
    \centering	
    \includegraphics[width=7.5cm]{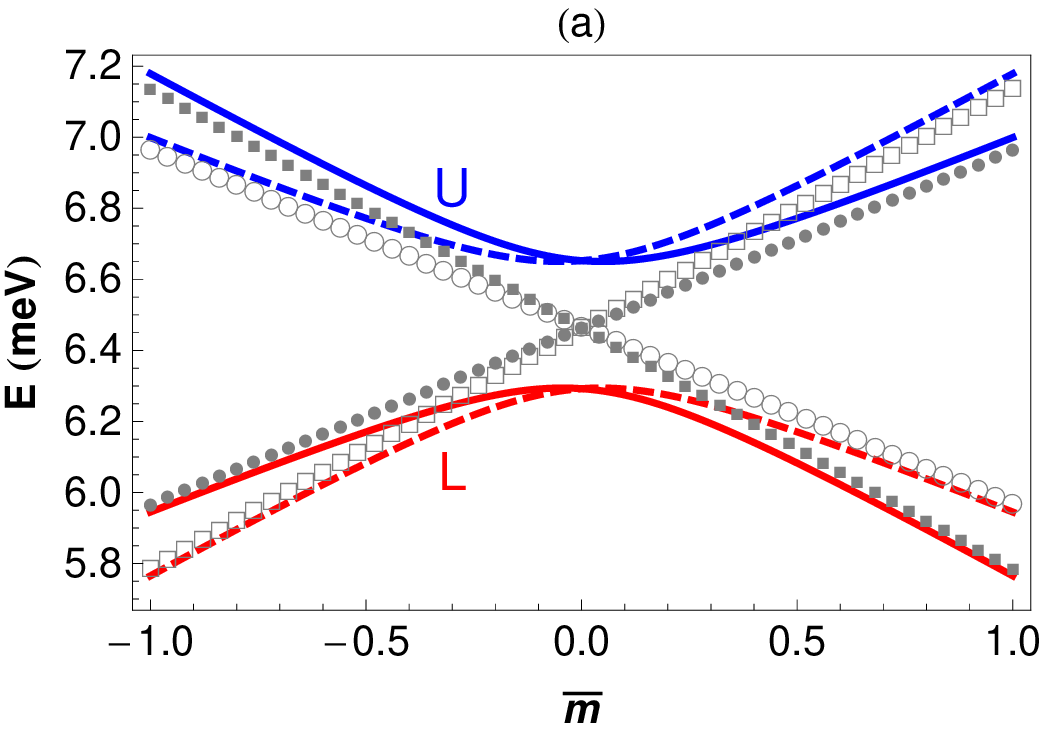}
    \includegraphics[width=7.5cm]{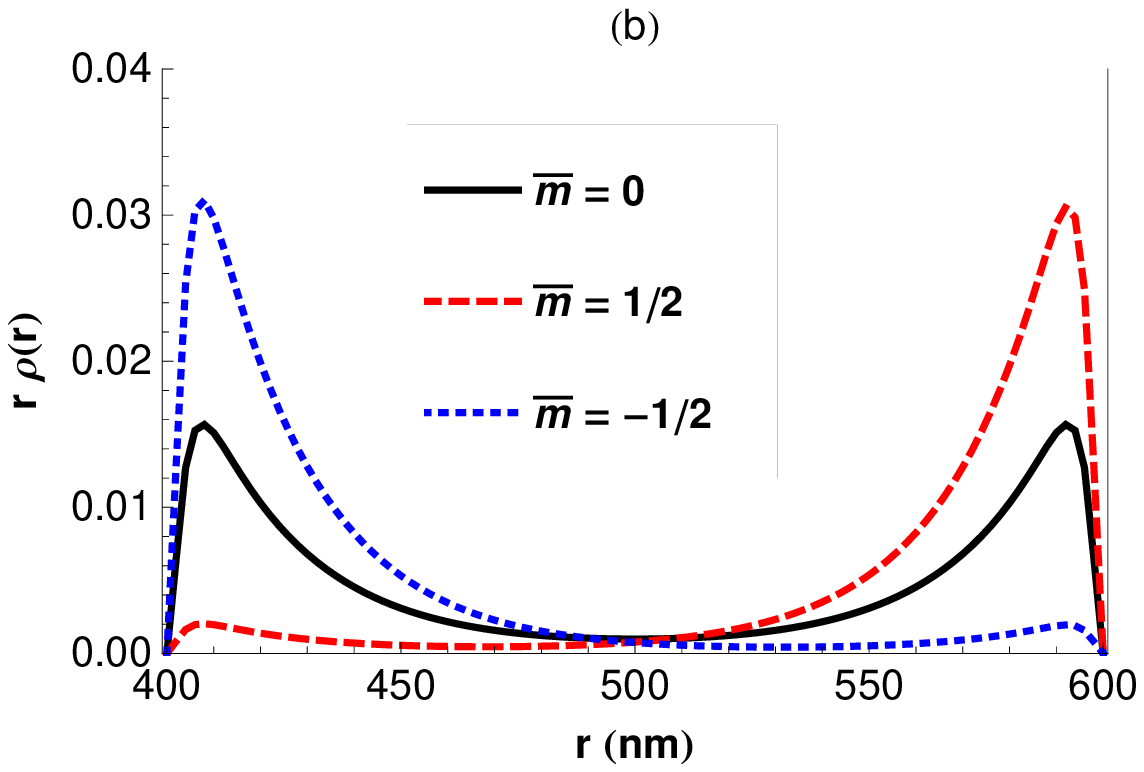}
    \caption{(Color online) (a) Upper (U) and lower (L) bands of the edge states for a ring with inverted 
      Dirac mass $M=-10$~meV, radius $a=500$~nm and width $w=200$~nm 
      for spin up and down blocks with solid ($\tau= 1$) and dashed ($\tau=- 1$) lines.
      Squares and circles represent the internal and external boundary (helical) edge states calculated independently, 
      filled for $\tau= 1$ and empty for $\tau=- 1$, which would result from 
      a very large width $w$ where no minigap develops.
      They could also describe a hole and disk geometry.
      (b) Radial probability density $\rho(r)=|\Psi(r)|^2$ of the edge states of the upper band with $\tau=1$, 
      for $\bar{m}=0$ and $\bar{m}=\pm\frac{1}{2}$. 
    }
    \label{fig:edge}
  \end{figure}

 \begin{figure}[tb]
    \centering	
    \includegraphics[width=7.5cm]{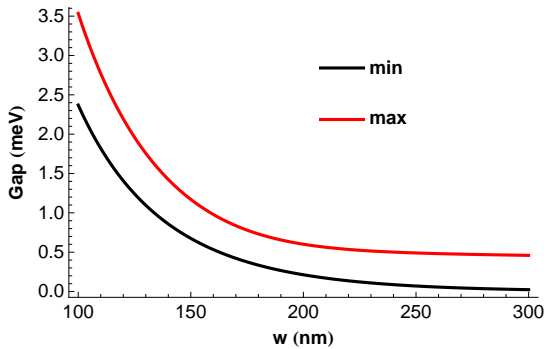}
    \caption{(Color online) Minimal and maximal values of the minigap  (it is a periodic function of the magnetic flux) 
      for a ring with radius $a=500$~nm as a function of $w$ between $100$~nm and $300$~nm.   
    }
    \label{fig:minigap}
  \end{figure}

  \begin{figure}[tb]
    \centering	
    \includegraphics[width=7.5cm]{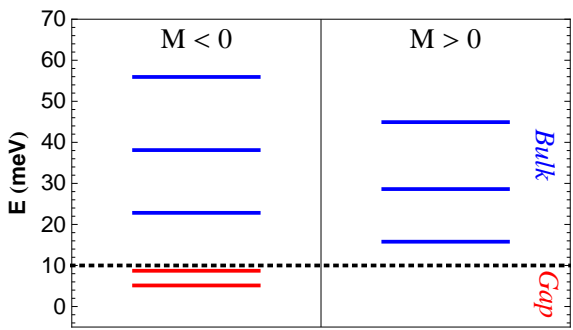}
    \caption{(Color online) Low-lying bound states corresponding to $\bar{m}=\frac{1}{2}$, $\tau=1$, a Dirac rest mass of $M=\pm10$~meV, ring radius $a=500$ nm and width $w=100$ nm.
      The change of sign of the mass term is accompanied by the disappearence of the edge states and 
        by a non-trivial shift of all eigenenergies.
    }
    \label{fig:levels}
  \end{figure}

  \subsection{Boundary conditions and secular equation}
  %
  %
  In order to obtain the energy levels and bound states of the ring, we need to solve the boundary problem.
  We assume a ring of radius $a$ and width $w$ and impose vanishing boundary conditions at $r=a\pm \frac{w}{2}$
  \begin{eqnarray}
    \chi_1^{m,\tau}\left(a \pm \frac{w}{2}\right)&=&0   \label{eq:boundary1}\\
    \chi_2^{m,\tau}\left(a \pm \frac{w}{2}\right)&=&0.
    \label{eq:boundary2}
  \end{eqnarray}
  Details of this calculation can be found in Appendix~\ref{app:spinors}. 
  Here we just note that the 
  four former conditions plus the normalization of the wave function fix $\vec{h}$ and lead to a 
  transcendental secular equation for $E$. 
  We here remark in connection with rings in graphene \cite{recher2007}, that
  the inclusion of quadratic terms allows for a formulation of vanishing boundary conditions which is not allowed
  in the Dirac equation (only linear terms in $k$) of graphene as the wave function would vanish identically. 
  There, a confinement on a circle can be formulated with an infinite mass boundary condition \cite{berry1987, recher2007}.

  For the bulk-propagating region we obtain the secular equation described by Eq.~(\ref{eq:sec_bulk}), 
  while for the edge states we obtain Eq.~(\ref{eq:sec_edge}).
  For a fixed $m$ and $\tau$, the secular equation is numerically solved.
  In the bulk-propagating region we obtain
  the eigenvalues $E_{n,m,\tau}$ where $n=\pm2$, $\pm3$, $\dots$ represents a radial quantum number, 
  positive for conduction band modes and negative for valence band modes.
  The corresponding eigenstates are given by the spinors $\Psi_{n,m,\tau}$ or in ket notation $|n,m,\tau\rangle$.
  We can think of the edge states, when system parameters allow their existence, 
  as the system's eigenstates with $n=\pm1$.

  \begin{figure}[tb]
    \centering	
    \includegraphics[width=7.5cm]{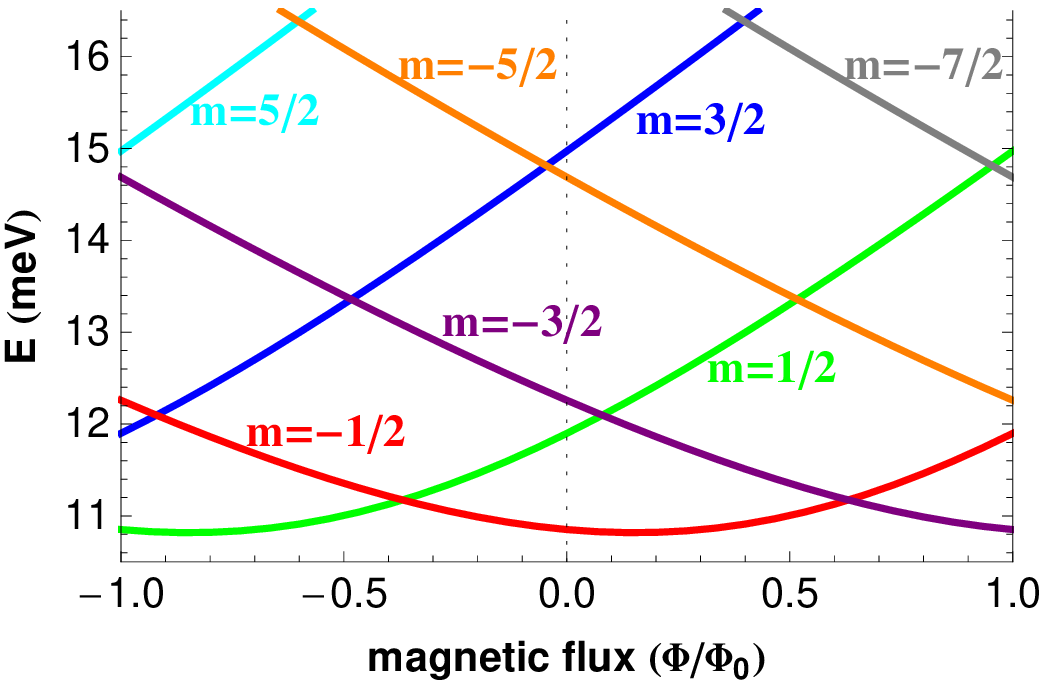}
    \includegraphics[width=7.5cm]{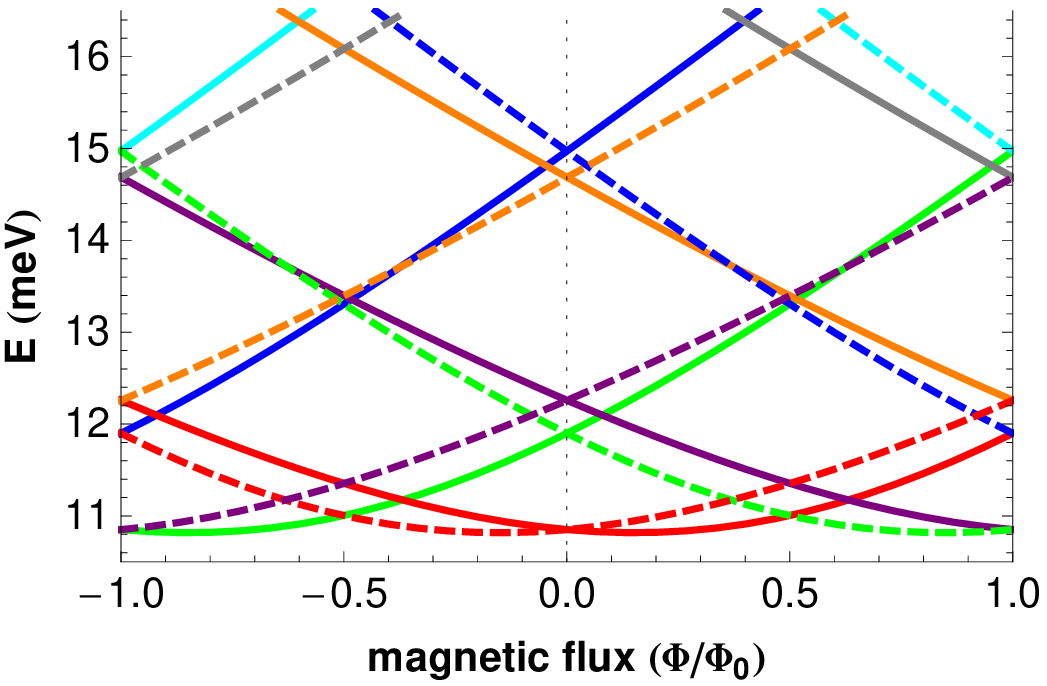}
    \caption{(Color online)  Dispersion curves for the first bulk-conducting mode $n=1$, as a function of the 
      magnetic flux for a ring in the inverted regime ($M=-10$~meV) of radius $a=100$~nm and 
      width $w=75$~nm (note that this mode originates from the edge states 
      pushed into the bulk conduction band by their overlap).
      States for $m=\pm1/2$, $\pm3/2$ and $\pm5/2$ are shown, in (a) for the block $\tau=1$ only, 
      and in (b) for both blocks $\tau=\pm1$ with continuous and dashed lines, respectively. 
    }
    \label{fig:dis_bulk}
  \end{figure}
   \begin{figure}[tb]
    \centering	
    \includegraphics[width=7.5cm]{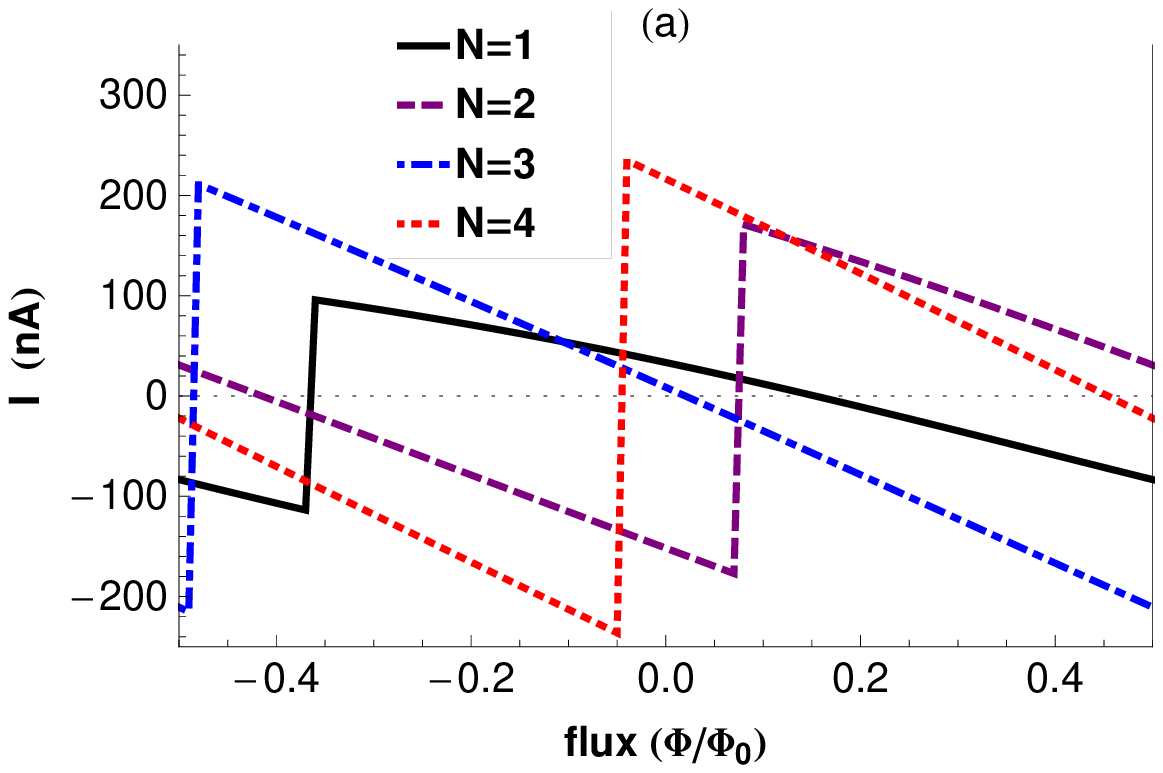}
    \includegraphics[width=7.5cm]{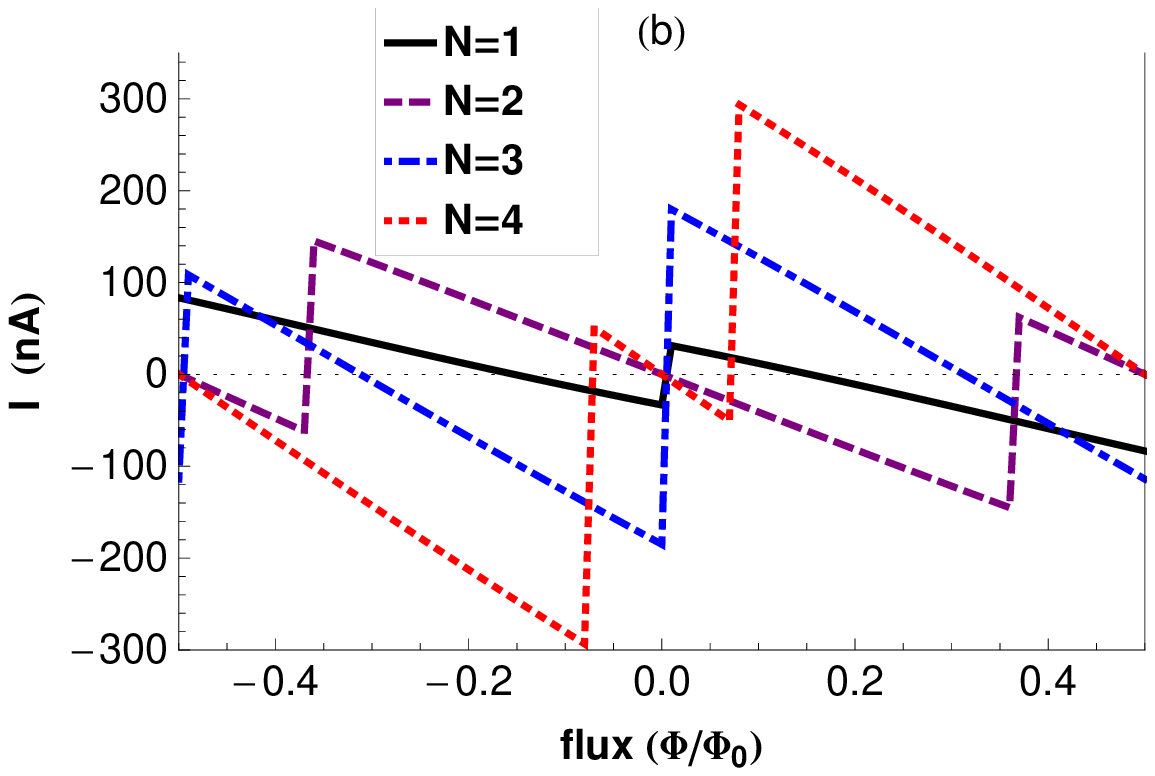}
    \caption{(Color online)  Spin-dependent persistent currents calculated for $N=1$, $2$, $3$ and $4$ electrons 
      in the ring's first bulk-propagating mode as a function of the 
      magnetic flux for a ring of radius $a=100$~nm and width $w=75$~nm.
      In (a) only the spin block $\tau=1$ is occupied, while in (b) electrons can occupy both spin-blocks.
    }
    \label{fig:pers_cur_bulk}
  \end{figure}

  \section{Bound states and persistent current of the ring}
  When the mass parameter $M$ is negative (the topological non-trivial regime),
  our model accepts solutions in the (inverted) gap region $E\in\left (M,-M\right)$.
  This solutions are exponentially localized near the external and internal boundaries 
  $r=a\pm\frac{w}{2}$ respectively, and are addressed as edge states.
  In Fig.~\ref{fig:edge}(a) we show the edge state dispersion for a relatively large and wide ring of radius $a=500$~nm 
  and width $w=200$~nm calculated with Eq.~(\ref{eq:sec_edge}) for a boundary condition including both ring edges.
  The physical content of this picture can be undestood remembering that for a given magnetic flux $\Phi$, 
  the eigenstates of the ring are found at $\bar{m}= m +\Phi/\Phi_0$ with $m$ an half-odd integer.
  It means that with $\Phi$ we can move the eigenstates along the dispersion curves in Fig.~\ref{fig:edge}(a), 
  always having two subsequent states at unitarily spaced intervals 
  on the $\bar{m}$-axis, i.e. for $\Phi=0$ we have states at $\bar{m}=\pm1/2$.
  The curves with squares and circles represent the internal and external 
  edge states respectively, as calculated from the decoupled boundary conditions 
  with Eqs.~(\ref{eq:sec_int_edge}) and (\ref{eq:sec_ext_edge}).
  We can also think of them as originated from edge states of a disk (external one) 
  and of a hole in a plane (internal one).
  Their dispersions are different because they are centered on two different effective radii 
  (approximately at $r=a\pm\frac{w}{2}$). 
  We note that their slope (which is an energy) is in fact approximately given~\cite{jiang2009}
  by $\frac{A}{r}\sqrt{\frac{B^2-D^2}{B^2}}$. 
  This emerges naturally from the fact that the radial wave function has to close on itself in a 
  roundtrip along the circumference.

  When the ring width is small enough to permit a finite overlap of the helical edge states, 
  internal and external edge states get mixed.
  Consequently, as shown in Fig.~\ref{fig:edge}(a), the dispersion curve is modified into the anticrossing 
  upper and lower bands separated by a minigap, 
  as it was also predicted for a straight TI ribbon of finite width~\cite{zhou2008}.
  In Fig.~\ref{fig:edge}(b), we show the radial probability density of the edge state spinors 
  of states at different points along the upper band (U) for $\tau=1$.
  This picture gives us an understanding of the localization properties of the edge states near the minigap. 
  The state at $\bar{m}=0$ (at the anticrossing) is equally shared between the internal and external edges. 
  Away from this anticrossing point (we plot the states with $\bar{m}=\pm\frac{1}{2}$), 
  where the dispersion curve resembles that of the uncoupled edges, the edge states are strongly 
  localized near one or the other boundary.
  This means that we are able to drag an electron from one edge of the ring to the other by tuning $\Phi$.
  Conversely, for the lower (L) band $\bar{m}=\pm\frac{1}{2}$ states exhibit opposite localization 
  properties as compared to the upper band. 
  The implications of this tunability for quantum information processing will be discussed in Section V.

  For certain values of the flux (when a state is in correspondence with $\bar{m}=0$), 
  as expected, the minigap exponentially shrinks with increasing $w$ (see Fig.~\ref{fig:minigap}). 
  However in a fully confined system, as our ring, the discreteness of 
  levels can also lead to a saturation of the minigap at a finite value.  
  For sufficiently small $w$ the edge modes can be pushed into the bulk-region, losing their edge-character 
  and becoming the first of the bulk-propagating modes.
  In this case, the minigap is comparable or larger than the bulk gap itself (i.e. larger than $2|M|$).

  The presence of the edge states with energy in the bulk gap depends on the negativity of the
  Dirac rest mass $M$, while for positive $M$ only bulk states exist.
  But the change of the mass sign has also effects on the bulk energy levels as can be seen in Fig.~\ref{fig:levels}, 
  which shows the lowest lying eigenergies with $\bar{m}=\frac{1}{2}$ and $\tau=1$ for a ring of 
  radius $a=500$~nm, width $w=100$~nm and for $M=\pm10$~meV.   
  Therefore the presence of the topological insulator phase  in confined systems can also be deduced from bulk-propagating properties as measured by optical spectroscopy or tunneling transport experiments through the ring.

  In Fig.~\ref{fig:dis_bulk}, we show the dispersion curve as a function of the 
  magnetic flux for the first bulk-conducting mode for a relatively small 
  and narrow ring of radius $a=100$~nm and width $w=75$~nm.
  We select a ring of small radius in order to enhance the magnitude of the persistent current, 
  which is proportional to the derivative of the dispersion curves with respect to the magnetic flux, 
  which grows with decreasing $a$.   
  The small confinement length $w$ leads, instead, to well separated eigenvalues corresponding to different 
  radial quantum numbers $n$.
  The energy level spacing, between energy eigenvalues of different radial quantum number $n$ 
  (see Fig~\ref{fig:eigen_bulk}), scales, in fact, with $1/w$. 
  In Fig.~\ref{fig:dis_bulk}(a) we show the states of the spin block $\tau=1$ with $m=\pm1/2$, $\pm3/2$ and $\pm5/2$.
  As can be seen, also in the absence of the magnetic field, effective time reversal symmetry $(m\rightarrow -m)$
  is broken within a single block.
  The presence of the second spin-block restores time reversal symmetry when $\Phi=0$, and 
  states with the same $n$, and  opposite $\tau$ and $m$ are time reversal (or Kramer) partners.
  The dispersion curve is periodic in $\Phi$ with periodicity given by the flux quantum $\Phi_0$.

  The persistent current at zero temperature is given by 
  \begin{equation}
    I=- \sum_{n,m,\tau} \frac{\partial E_{n,m,\tau}}{\partial \Phi},
  \end{equation}
  where the sum runs over all occcupied states.
  In Fig.~\ref{fig:pers_cur_bulk} we show the persistent current for a ring with the same 
  parameters as used for Fig.~\ref{fig:dis_bulk}. 
  We consider an electron occupation number $N$ up to $4$ occupying the first radial mode with positive energy,
  subtracting the current contribution coming from all states below.
  In Fig~\ref{fig:pers_cur_bulk}(a) we consider injection of spin-polarized electrons corresponding to $\tau=1$.
  As a consequence, the persistent current is finite even in the absence of a magnetic flux, and its sign directly reflects the spin polarization.
  While in Fig~\ref{fig:pers_cur_bulk}(b), both blocks can accommodate electrons 
  and, accordingly, the current vanishes at zero magnetic field, respecting the time reversal symmetry of the model. 
  However, we note that the spin current can be finite at $\Phi=0$.	
  The currents are periodic in $\Phi$ with the same periodicity as the dispersion curves in Fig.~\ref{fig:dis_bulk}, 
  and show the presence of kinks whenever different bands cross and electrons are forced to move from one to 
  the other in order to maintain the groundstate of the system.

  The persistent currents in Fig.~\ref{fig:pers_cur_bulk} are calculated for an edge state ($n=1$) pushed into the 
  bulk conduction band by the overlap between internal and external edge states.
  But generally, a similar behavior is observed for other bulk states of the ring ($|n|>1$, not shown), 
  where a spin-selective behavior is not as obvious as in the contribution from the helical edge states. 
  The reason for this phenomenon is that the Dirac mass $M(k)$ breaks the symplectic 
  symmetry ${\vec p}\rightarrow -{\vec p}$ and $\sigma\rightarrow -\sigma$ in a single block. 
  In graphene rings with a smooth ring boundary the same symmetry is broken and leads to a 
  valley-dependent persistent current~\cite{recher2007}.
  This physics in principle would 
  allow to use rings in topological insulators as spin-detectors where the direction 
  of spin is mapped to the direction of the persistent current.
  Indeed currents shown in Fig~\ref{fig:pers_cur_bulk}(a) and (b) are well in the range 
  accessible to experiments~\cite{mailly1993}.
  The persistent current has also been calculated for a disk in an homogeneous magnetic field \cite{chang2010}.
  However, the topology of that system is qualitatively different from a ring.

  %
  \begin{figure}
    \includegraphics[width=7.5cm]{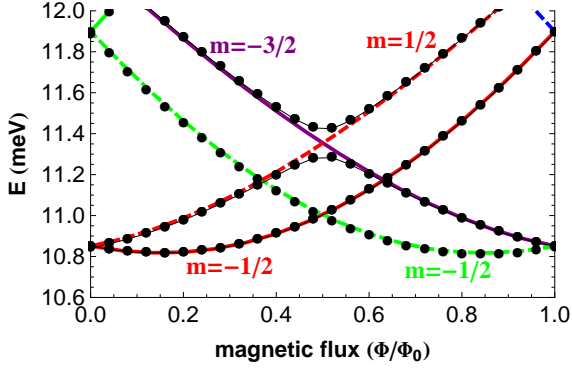}
    \caption{(Color online)
      Details of the dispersion curve of the first radial mode in the presence of the Rashba term 
      ($\alpha_0=20$~meV~nm) for a ring of radius $a=100$~nm and width $w=75$~nm. 
      Lines are dispersion curves without the Rashba term and points after its inclusion 
      (a thin line is used to link the points to emphasize the spin-splitted bands).
      The Rashba term induces a splitting at the crossings of $|1,-\frac{3}{2}, 1\rangle$ with $|1,\frac{1}{2}, 
      -1\rangle$, and of $|1,-\frac{1}{2},1\rangle$ with $|1,\frac{3}{2},-1\rangle$ (not visible in this range of $\Phi$).}
    \label{fig:rashba}
  \end{figure}

  \section{Rashba spin-orbit interaction}
  We consider the inclusion of a Rashba spin-orbit interaction as derived in Ref.~\onlinecite{rothe2010}, retaining 
  only terms to first order in $k$, which gives the most dominant contribution
  \begin{eqnarray}
    \tilde{H}_R = \alpha_0 \frac{(\sigma_0+\sigma_z)}{2} \left( \tau_y k_x -\tau_x k_y \right).
  \end{eqnarray} 
  We use cylindrical coordinates and transform it into the symmetrized basis using Eq.~(\ref{eq:trans}). We obtain 
  \begin{eqnarray}
    H_R = 
    i \alpha_0 \frac{(\sigma_y-i \sigma_x \tau_z)}{2} e^{-i \tau_z \theta} \left( \tau_y \partial_r -
      \tau_x \frac{\partial_\theta}{r} \right),
  \end{eqnarray} 
  where $\alpha_{0}$ is the Rashba coefficient.
  In particular, the Rashba interaction in HgTe quantum well can be tuned with a gate voltage, so that the induced 
  spin-splitting gap varies from zero up to tens of meV~\cite{privat}.

  We note that $H_{R}$ does not commute with $H$ and $\hat{j}_z$, 
  and therefore will in general mix energy eigenstates of $H$.
  We can still find a conserved physical observable by noting that the only effect of $H_R$ is 
  to couple $|E_1,+\rangle$ with $|E_1,-\rangle$. 
  This motivates us to define the total angular momentum restricted to the $E_1$-band~\cite{note}
  \begin{equation}
    \hat{j}_z^{(E_1)} = \hat{l}_z^{(E_1)} + \hat{S}_z^{(E_1)},
  \end{equation}
  where the operators are the same as described in Table~(\ref{tab:1}), but their action 
  is now restricted to states of the $E_1$ band only. 
  This can be obtained by applying the projector $P^{(E_1)}=\frac{\sigma_0+\tau_z\sigma_z}{2}$ 
  to the operators defined in Table~\ref{tab:1}, so that $\hat{l}_z^{(E_1)}= \hat{l}_z P^{(E_1)}$ 
  and $\hat{S}_z^{(E_1)}= \hat{S}_z  P^{(E_1)}$.
  It is clear that the Rashba term $H_R$ must conserve $\hat{j}_z^{(E_1)}$ and indeed 
  $[H_R,  \hat{j}_z^{(E_1)}] = 0$.
  On the other hand the spinors $\Psi_{n,m,\tau}$ are eigenstates of $H$ and 
  eigenstates of $\hat{j}_z^{(E_1)}$ with eigenvalue $m+\tau$   
  \begin{equation} 
    \hat{j}_z^{(E_1)} \Psi_{n,m,\tau} = \hbar \left(m+\tau\right) \Psi_{n,m,\tau}.
  \end{equation}
  Therefore, $H_R$ can only couple 
  eigenstates of $H$ from the blocks $\tau=1$ and $\tau=-1$ with  
  \begin{equation}
    m_{(\tau=-1)}-m_{(\tau=1)}=2.
    \label{eq:rule_rashba}
  \end{equation}

  The conservation of the ``$E_1$-band'' total angular momentum imposes that the only non-trivial matrix 
  elements exist between the spinors $|n, m, 1\rangle$ and $|n', m+2, -1\rangle$, 
  with $n$ and $n'$ two quantum numbers of the radial quantization, 
  and $m$ the quantum number of the total angular momentum.    
  In the following we will assume the eigenvalues of two different radial eigenstates ($n\ne n'$) to be 
  sufficiently distant in energy, and the Rashba term sufficiently small, such as to make the mixing induced 
  by spin-orbit negligible.
  Actually the strength of the Rashba term can be tuned in topological insulators, 
  while the distance between neighbouring radial eigenstates increases for smaller $w$, 
  so that we can always restrict ourselves in a parameter range which makes the previous assumption valid.

  The relevant matrix element of $H_R$ is  
  \begin{eqnarray}
    H_{R(n,m)} = \langle n, m+2, -1| H_R |n, m, 1\rangle = \nonumber\\
    = i \alpha_0 \int dr r \chi_2^{(\bar{m}+2,-1)*}\left(\partial_r - \frac{\bar{m}+\frac{1}{2}}{r}\right) \chi_1^{(\bar{m},+1)},
  \end{eqnarray} 
  which, using the Bessel function's property Eq.~(\ref{eq:prop_bessel}), can be put in the form
  \begin{multline}
    H_{R(n,m)}\hspace{-0.1cm}\\=\hspace{-0.1cm} -i\alpha_0\hspace{-0.1cm} \int\hspace{-0.1cm} dr r \hspace{-0.1cm}\left[ \vec{h} \mathbb{R} \vec{F}_{\bar{m}+\frac{3}{2}} \right]_{\hspace{-0.05cm}|n,m\hspace{-0.02cm}+\hspace{-0.02cm}2,\hspace{-0.02cm}-\hspace{-0.02cm}1\rangle}^*\hspace{-0.15cm} \left[\vec{h}\mathbb{K} \vec{F}_{\bar{m}+\frac{1}{2}}   \right]_{\hspace{-0.05cm}|n,m,1\rangle},
  \end{multline}
  with $\mathbb{K}={\rm diag}\left(K_1,\eta K_1,K_2,-K_2 \right)$,
  and where the coefficients $\vec{h}$, $\mathbb{R}$ and $\mathbb{K}$, in squared parenthesis,
  are calculated for the indicated spinors $|n, m, \tau\rangle$.  
  This expression is valid for both bulk and edge states, for which $\eta=\pm1$, respectively.

  In Fig.~\ref{fig:rashba}, we show the effect of a Rashba term with $\alpha_0=20$~meV$\times$nm on the dispersion 
  relation of the first bulk modes of a ring with the same parameters as in Fig.~\ref{fig:dis_bulk}.
  We note that the most significant signature is the opening of a splitting with an anticrossing behavior between the
  states $|1,-\frac{3}{2},1\rangle$ and $|1,\frac{1}{2},-1\rangle$, and between their time reversed ones
  $|1,-\frac{1}{2},1\rangle$ and $|1,\frac{3}{2},-1\rangle$ (not shown).

    \begin{figure}[tb]
    \centering	
    \includegraphics[width=7.cm]{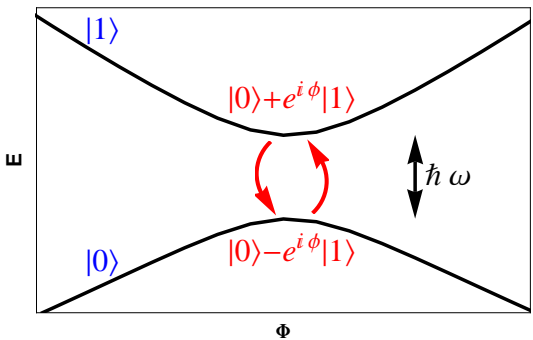}
    \caption{(Color online) Schematic view of a flux tunable two-level system.
      $|0\rangle$ and  $|1\rangle$ states refer to internal and external edge states, 
      when the splitting $h\omega$ is the minigap. 
      Another definition is $|1\rangle=|\tau=+1\rangle$ and $|0\rangle=|\tau=-1\rangle$, 
      when we are dealing with the spin degree of freedom and the energy splitting is due 
      to the Rashba interaction.
    }
    \label{fig:qbit}
  \end{figure}

  \section{Coherent manipulation of spin and edge degrees of freedom}
  An AB-ring in a two-dimensional topological insulator can be used to process quantum information. 
  Our idea is to use the magnetic flux to manipulate a two-level system of states $|1\rangle$ and $|0\rangle$, 
  which are good system eigenstates, 
  but which can be coupled by some interaction for specific values of the flux (see the scheme in Fig.~\ref{fig:qbit}).
  At resonance, corresponding to the operational magnetic flux $\Phi_c$, 
  the eigenstates of the system can be described by
  $|U\rangle=\left(|1\rangle+e^{i \phi} |0\rangle\right)/\sqrt{2}$ and $|L\rangle=\left(|1\rangle-e^{i \phi} |0\rangle\right)/\sqrt{2}$, 
  for the upper ($U$) and lower ($L$) band.

  We identify in our TI ring two pairs of such two-level systems.
  The first one, which is specific to a topological insulator ring, makes use of 
  the external ($|1\rangle$) and internal ($|0\rangle$) edge states, which realize a system similar to that of  
  a charge-qubit in a double quantum dot~\cite{hayashi2003}, but where charge separation is due to the nature of the edge states.
  A ring has actually an even number of Kramer's partner and therefore it is topologically trivial, 
  in the sense that scattering is allowed between the two different set of helical edge states 
  localized at the internal and at the external boundaries.
  However for sufficiently large $w$ a degree of the topological protection remains, 
  because the scattering is suppressed by their localization properties. 
  As shown in Fig.~\ref{fig:edge}(b), at zero flux, states of the same spin $\tau$ with $m=1/2$ and $-1/2$ 
  are strongly (exponentially) localized at the external and internal boundary respectively. 
  Let us imagine to inject a state of definite spin $\tau$  into the state $m=1/2$ 
  which is strongly localized at a single edge (see Fig.~\ref{fig:edge}) and then non-adiabatically 
  vary the flux from zero to the operational value $\Phi_c=-\Phi_0/2$, corresponding to $\bar{m}=0$.
  In this situation the eigenstates of the system are equally shared between the two edges and an energy 
  splitting $\hbar \omega$ (the minigap) arises, due to the inter-edge overlap.
  The original state will now be subjected to Rabi oscillations between the internal and 
  the external boundaries with angular velocity $\omega$.
  Read out of this two-level system could be performed by connecting separately, 
  with tunneling contacts, the internal and external boundary, or measuring the sign of the 
  flux generated by the persistent current (cf. Fig.~\ref{fig:pers_cur_bulk}).

  Another, more canonical choice, initially proposed in Ref.~\onlinecite{loss1998} 
  is to use the spin degree of freedom and to associate $|1\rangle=|+\rangle$ and $|0\rangle=|-\rangle$ 
  with $\pm$ referring to the upper and lower block of the TI 
  Hamiltonian with $\tau_z=\pm1$.
  This two-level system is indeed topologically protected in a ring for a sufficiently large $w$, 
  or in a disk or hole geometry, from time-reversal symmetric scattering, but still is subject to external 
  decoherence sources, like the hyperfine interaction with an underlying nuclear spin system. 
  The tunable Rashba term can offer a way to manipulate the spin-degree of freedom~\cite{debald2005}, 
  as implied by Fig.~\ref{fig:rashba}, where 
  a spin splitting gap ($\hbar \omega$) opens between the state $|+\rangle=|m, \tau_z=+1\rangle$ and  
  the state $|-\rangle=|m+2, \tau_z=-1\rangle$ around $\Phi=\Phi_0/2$.
  The spin state could be observed by detected the persistent current, 
  which is opposite for the two spin states (cf. Fig.~\ref{fig:pers_cur_bulk}).

  Both spin and edge (i.e. charge) two-level systems can be operated by the magnetic flux.
  For example, let us imagine to be initially in state $|1\rangle$, in a range of the flux value for which 
  it is a good eigenstate of the system (see Fig.~\ref{fig:qbit}).
  Swiftly turning the magnetic flux to the operational value $\Phi_c$, and keeping it there
  for a time $t$, will lead to the final rotated state described by  
  $\cos{(\omega t/2)}|1\rangle + i e^{-i\phi} \sin{(\omega t/2)} |0\rangle$.
  We can switch off the rotation by restoring the magnetic flux to its initial value.
  For the stability of the two-level system it is required that the dephasing and relaxation time are much larger 
  than the Rabi period $2\pi/\omega$.  
  For these two kinds of two-level systems (edge and spin), the splitting can be on the order of $1$~meV 
  for reasonable parameters $a$ and $w$, corresponding to a Rabi period of $4$~ps.

  One can also exploit the ability to control the localization properties of 
  an edge electron, that is, the possibility to move it from the internal to the external boundary by means of the flux,
  in a ring network.
  In such a system, spin qubits can be \emph{stored} in the edge states localized on the internal boundary, 
  the position in which they are protected from the interaction with the network.
  Bringing them to the external boundary, corresponds instead to put them in an \emph{active} position, 
  allowing for their measurements or for the interaction with qubits on other rings.

  \section{Conclusion}
  In the present paper we have solved the eigenstates of a topological insulator for a ring, 
  a disk and a hole structure, taking HgTe quantum wells as a practical example.
  We solved analytically the four-band model with both linear and quadratic terms, yielding a 
  transcendental secular equation which we solved numerically as a function of the magnetic flux threading the ring.
  We calculated persistent currents, which can be exploited to measure the spin state of the system.
  The effect of a Rashba spin-orbit term has also been taken into account, showing that its main effect is 
  the opening of a spin-splitting gap at specific values of the magnetic flux. 
  We proposed that, in a TI ring, spin and edge (charge) 
  degrees of freedom, where either up/down spins or internal/external edges are used as quantum basis states 
  (two-level systems), can be coherently processed.
  Rashba interaction can be used to manipulate spin, while a finite edge-state  overlap can 
  be exploited to manipulate the edge degree of freedom, whereas the AB-flux is used to introduce 
  coherent dynamics of the two-level system.

  \begin{acknowledgments}
  P. M. and P. R. acknowledge financial support from the DFG via the Emmy Noether program, and B. Trauzettel, 
  J. Budich, H. Buhmann and L.W. Molenkamp for helpful discussions.
  \end{acknowledgments}

  \appendix
  
  \section{Boundary conditions and secular equations \label{app:spinors}} 
  \begin{figure*}[tbh]
    \includegraphics[width=5.5cm]{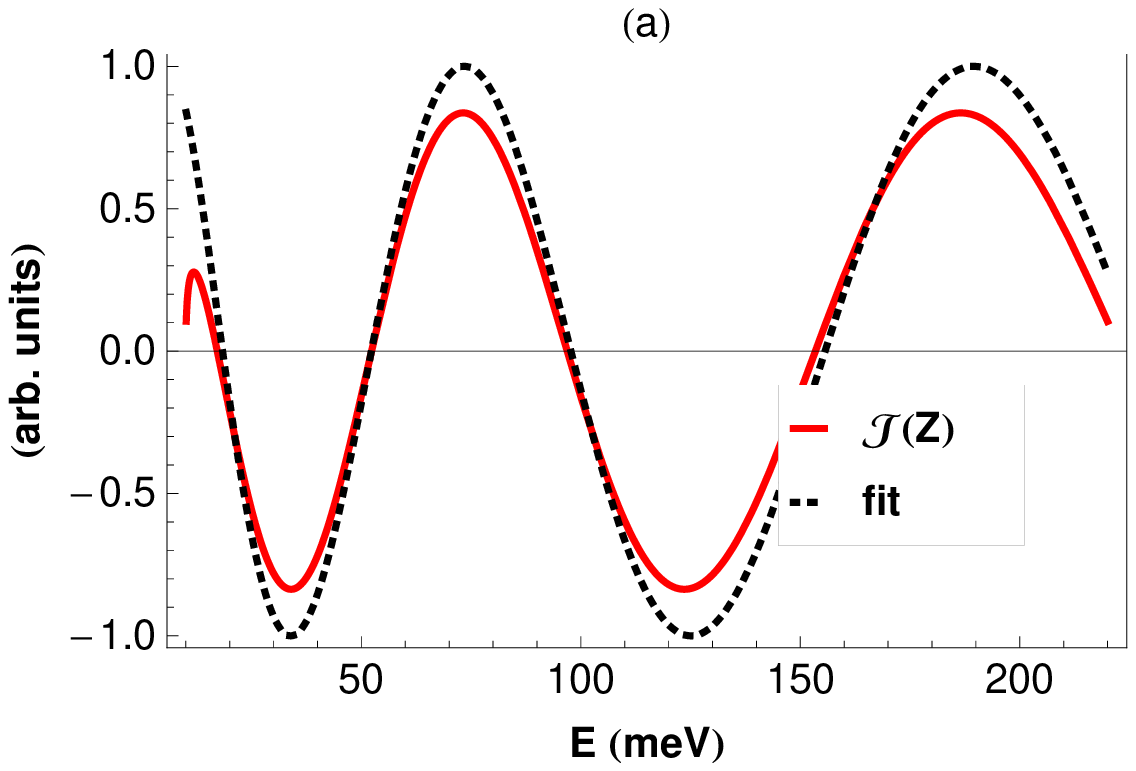}
    \includegraphics[width=5.5cm]{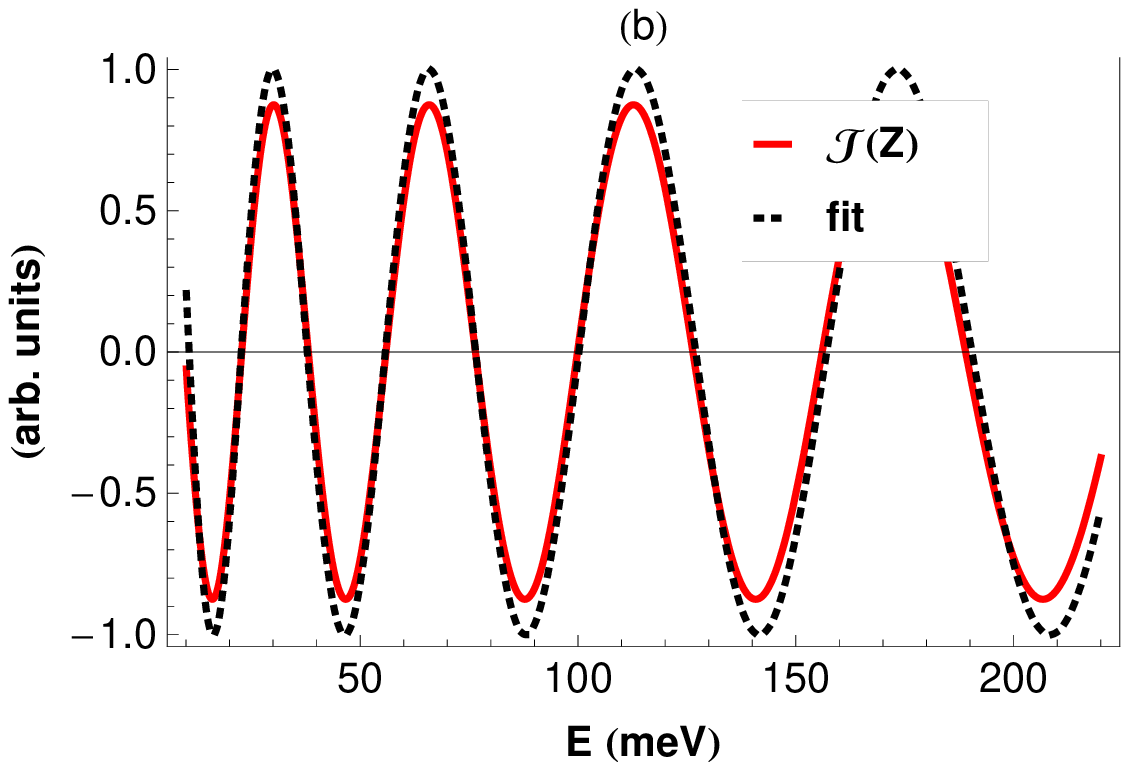}
    \includegraphics[width=5.5cm]{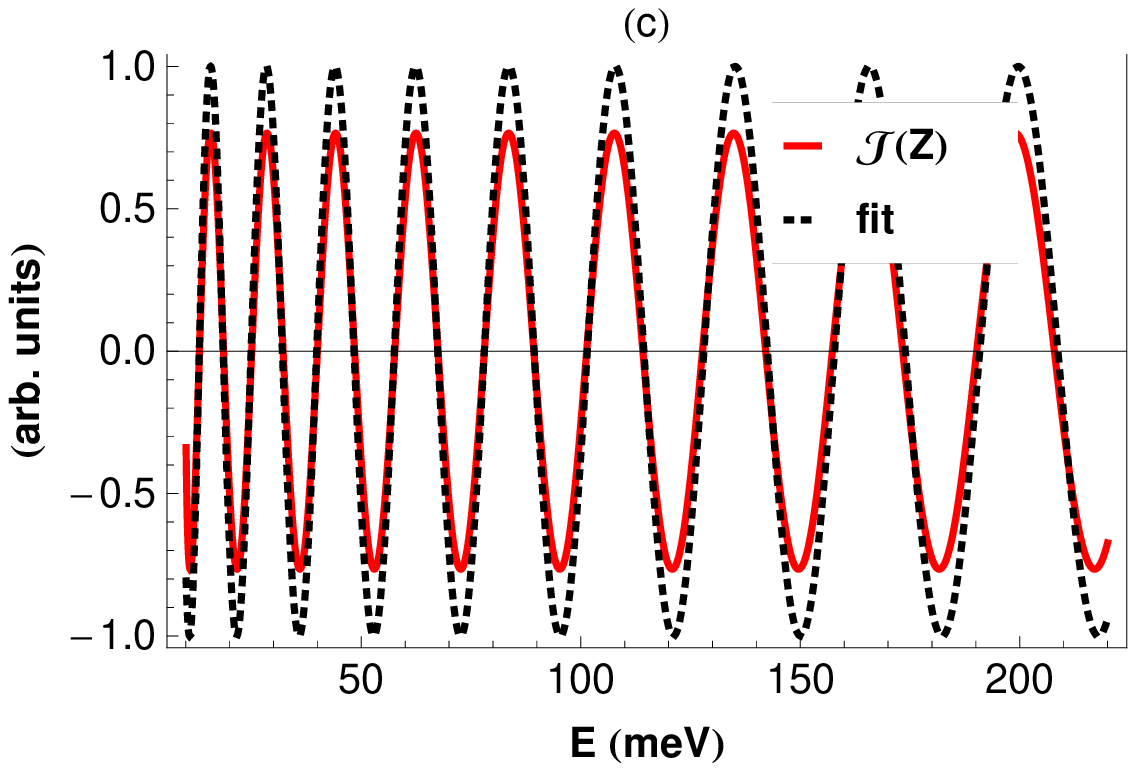}
    \caption{
      (Color online) $\Im{(z)}$ defined in Eq.~(\ref{eq:sec_bulk}), whose zeros give the eigenvalues of a ring with 
      radius $1000$~nm and width $50$ (a), $100$ (b) and $200$~nm (c).
      The dashed curve is an analytic fitting curve, described in the text.}
    \label{fig:eigen_bulk}
  \end{figure*}
  \subsection{Bulk-propagating modes}
  The bulk-propagating eigenstates of a TI ring, of total angular momentum $m$, can be written as 
  as a linear combination of the spinors of the four modes $K=\pm K_1$, and $\pm i K_2$ as 
  described by Eqs.~(\ref{eq:spinor}), (\ref{eq:comp1}) and (\ref{eq:comp2}).
  Now, vanishing boundary conditions (Eq.~(\ref{eq:boundary1}) and (\ref{eq:boundary2})) impose the spinor 
  components to vanish on the internal and external border and, toghether with the normalization condition, 
  fix both the four coefficients $h_n$ and determine a the secular equation for the system eigenvalues.
  We obtain the following secular equation 
  \begin{widetext}
    \begin{eqnarray}
      \Im{(Z)}&=&0 
      \label{eq:sec_bulk}\\
      Z&=& 
      \frac{ 
        A_2 
        (
        R_2 K_{-}(\xi_{2,1}) H_{+}(\xi_{1,1}) 
        +R_1 K_{+}(\xi_{2,1}) H_{-}(\xi_{1,1})
        ) 
        -A_1
        (
        R_2 K_{-}(\xi_{2,2})  H_{+}(\xi_{1,2}) 
        +R_1 K_{+}(\xi_{2,2})  H_{-}(\xi_{1,2})
        ) 
      }
      {
        R_2 A_-
        (
        K_{+}(\xi_{2,2}) H_{+}(\xi_{1,1})
        -K_{+}(\xi_{2,1}) H_{+}(\xi_{1,2}) 
        ) 
        -R_1 A_+
        (
        K_{-}(\xi_{2,2}) H_{-}(\xi_{1,1})
        -K_{-}(\xi_{2,1}) H_{-}(\xi_{1,2})
        )
      }\nonumber,
    \end{eqnarray}
  \end{widetext}
  where we adopted the following notations
  \begin{eqnarray*}
    \xi_{u,v}&=& K_u r_v; \hspace{1cm}  u,v=1,2\\
    r_1 &=& a - \frac{w}{2}; \hspace{1cm} r_{2}=a + \frac{w}{2}\\ 
    g_{\pm} &=& g_{m\pm \frac{1}{2}}; \hspace{1cm} g= H,K,I\\
    A_1 &=& I_+(\xi_{2,1}) K_-(\xi_{2,1}) + I_-(\xi_{2,1}) K_+(\xi_{2,1})\\
    A_2 &=& I_+(\xi_{2,2}) K_-(\xi_{2,2}) + I_-(\xi_{2,2}) K_+(\xi_{2,2})\\
    A_+ &=& I_+(\xi_{2,2}) K_+(\xi_{2,1}) - I_+(\xi_{2,1}) K_+(\xi_{2,2})\\
    A_- &=& I_-(\xi_{2,2}) K_-(\xi_{2,1}) - I_-(\xi_{2,1}) K_-(\xi_{2,2})
  \end{eqnarray*}
  Once the transcendental equation Eq.~(\ref{eq:sec_bulk}) is solved for the eigenenergy, the coefficients $\vec{h}$ 
  are obtained by solving the boundary conditions. 
  Note that these results depend on $\tau$ through the terms $R_{1}$ and $R_2$.

  For a given value of the quantum numbers $m$ and $\tau$, 
  the secular equation gives a series of eigenstates labeled by a radial quantum number $n$, 
  with $|n|$ corresponding to the number of nodes of the second component $\chi_2$.
  Generally with $M<0$ edge states have $n=\pm1$ and bulk propagating states are labeled by $n=\pm2$,$\pm3$,$\dots$, 
  where $\pm$ are for conduction and valence band states, respectively.
  When $M>0$ no edge states are present and bulk-propagating modes are labelled by  $n=\pm1$,$\pm2$,$\dots$.
  As an example, in Fig.~\ref{fig:eigen_bulk}, we show the numerical 
  solution for $\bar{m}=0$ of a TI ring in the inverterd regime ($M=-10$~meV) of radius 
  $a=1000$~nm and width $w=50$, $100$ and $200$~nm.
  The red curve corresponds to $\Im{(Z)}$ in Eq.~(\ref{eq:sec_bulk}) and its zeros define the 
  system eigenvalues.
  Note that the first zero, for the case with $w=50$~nm, corresponds to an edge state pushed into the bulk conduction 
  band by the overlap between internal and external edges (and it is therefore the $n=1$ state).
  The dotted lines correspond to the analytic fit $\sin{\left[K_1 \left(w-K_2^{-1}\right)\right]}$ 
  where $K_{1,2}$ are calculated at the energy $E+M\frac{\ell}{w}$, with the parameter $\ell\approx 50$~nm.
  As can be seen, this fitting curve well reproduces the spacing between the eigenvalues of the system. 
  The solutions are therefore determined by the energy $E$ for which $K_1 (w-K_2^{-1}) = n \pi$.
  With different values of the mass, we need to adequately fit a new $\ell$, which appears 
  to decrease with $M$.
  For $M=5$ and $15$~meV the best fit is given by $\ell=90$ and $\ell=37$~nm, respectively.

\begin{figure*}[tbh]
    \includegraphics[width=7.cm]{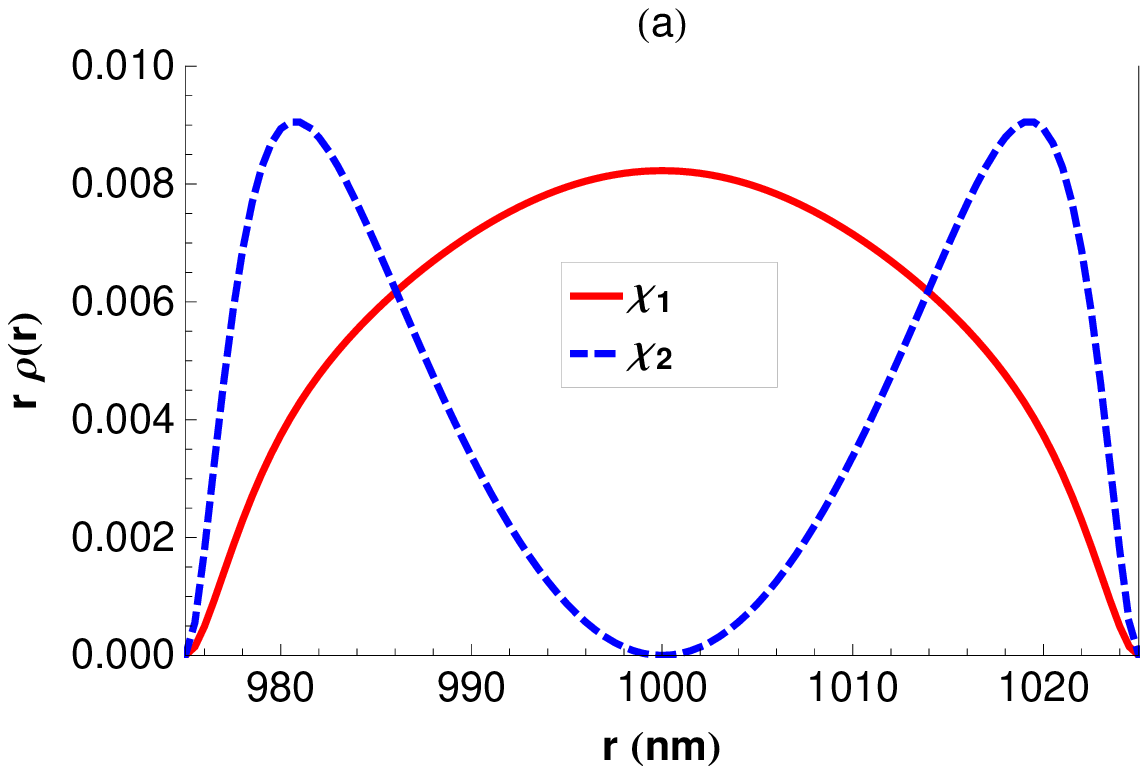}
    \includegraphics[width=7.cm]{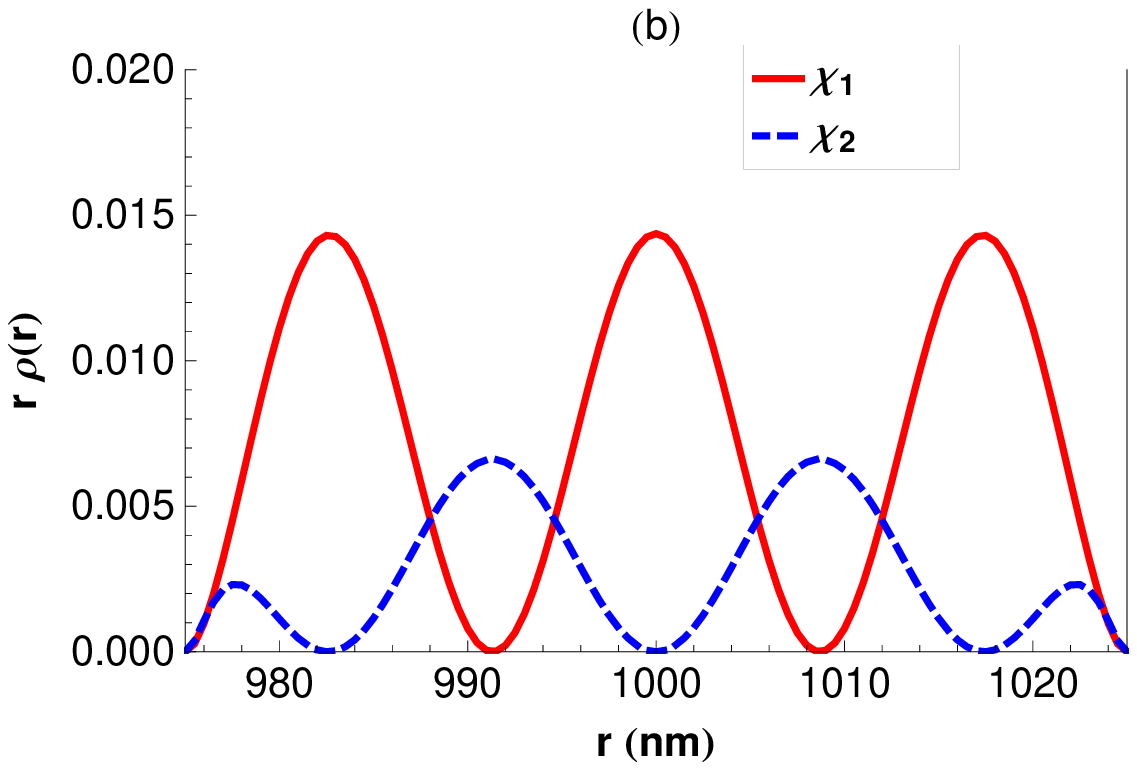}
    \caption{
      (Color online) Radial probability associated with the first and third mode for a ring with $a=1000$~nm and 
      $w=50$~nm and with negative Dirac rest mass $M=-10$~meV.} 
    \label{fig:spinors}
  \end{figure*}

  In Fig.~\ref{fig:spinors}(a) and (b) we show the $n=1$ and $n=3$ radial modes of the 
  system described in Fig.~\ref{fig:eigen_bulk}(a), corresponding to an energy of $17.2$~meV and $96.8$~meV, respectively. 

  \subsection{Edge states}
  In the gap spectral regions of a TI ring, we find evanescent solutions (i.e. helical edge states) if $M<0$.
  For a set of quantum numbers $m$ and $\tau$, Eqs.~(\ref{eq:spinor}), (\ref{eq:comp1}) and 
  (\ref{eq:comp2}), togheter with the definition of Eq.~(\ref{eq:edge}), describe the two spinor components.
  Now, solving vanishing boundary conditions (Eq.~(\ref{eq:boundary1}) and (\ref{eq:boundary2})) and 
  the normalization of the wavefunction, we are led to the secular equation
  \begin{widetext}
    \begin{eqnarray}
      Q &=& Z \label{eq:sec_edge}\\
      Q &=& \frac{A_2 \left( R_2 K_{-}(\xi_{2,1}) I_{+}(\xi_{1,1}) + R_1 K_{+}(\xi_{1,1}) I_{-}(\xi_{1,1})  \right) 
        - A_1 \left( R_2 K_{-}(\xi_{2,2} I_{+}(\xi_{1,2}) + R_1 K_{+}(\xi_{1,2}) I_{-}(\xi_{1,2} ) \right)}
      {  R_2 A_- \left(K_{+}(\xi_{2,2}) I_{+}(\xi_{1,1}) -  K_{+}(\xi_{2,1}) I_{+}(\xi_{1,2}) \right)
        -R_1 A_+ \left(K_{-}(\xi_{2,2}) I_{-}(\xi_{1,1}) -  K_{-}(\xi_{2,1}) I_{-}(\xi_{1,2}) \right)}\nonumber\\
      Z &=& \frac{A_2 \left(R_2  K_{+}(\xi_{1,1}) K_{-}(\xi_{2,1}) - R_1  K_{-}(\xi_{1,1}) K_{+}(\xi_{2,1}) \right)
        -A_1 \left(R_2  K_{+}(\xi_{1,2}) K_{-}(\xi_{2,2}) - R_1  K_{-}(\xi_{1,2}) K_{+}(\xi_{2,2}) \right)}
      {  R_2 A_- \left( K_{+}(\xi_{1,1}) K_{+}(\xi_{2,2}) - K_{+}(\xi_{1,2}) K_{+}(\xi_{2,1}) \right)
        + R_1 A_+ \left( K_{-}(\xi_{1,1}) K_{-}(\xi_{2,2}) - K_{-}(\xi_{1,2}) K_{-}(\xi_{2,1}) \right)
      }.\nonumber 
    \end{eqnarray} 
  \end{widetext}
  This equation fixes the eigenenergies, while the corresponding spinor coefficients $\vec{h}$ can be 
  obtained by solving, for a fixed energy, the boundary conditions.

  \subsubsection{Isolated boundaries: disk and hole}
  If we consider $w$ sufficiently large so as to effectively isolate the two boundaries, 
  or just consider a disk or hole geometry, 
  we can independently solve the internal ($-$ sign in Eqs.~(\ref{eq:boundary1}) and (\ref{eq:boundary2})) and external 
  ($+$ sign in Eqs.~(\ref{eq:boundary1}) and (\ref{eq:boundary2})) boundary conditions.
  The internal boundary conditions (hole geometry), for themselves, are compatible with $h_2=h_4=0$ and give
  \begin{eqnarray}
    I_{+}(\xi_{1,2}) + h_3 I_{+}(\xi_{2,2}) &=&0\\
    R_1 I_{-}(\xi_{2,1}) + h_3 R_2 I_{-}(\xi_{2,2})&=&0,
  \end{eqnarray}
  which results in the secular equation
  \begin{eqnarray}
    \frac{I_{+}(\xi_{1,2})}{I_{+}(\xi_{2,2})} =   \frac{R_1}{R_2}  \frac{I_{-}(\xi_{1,2})}{I_{-}(\xi_{2,2})}.
    \label{eq:sec_ext_edge}
  \end{eqnarray}
  For a given energy, the coefficients $h_1$ and $h_3$ are then obtained with the same boundary conditions. 
  The problem of holes in a TI has also been addressed in Ref.~\onlinecite{shan2010}.
  
  For the external boundary (disk geometry), consistent with $h_1=h_3=0$, we find
  \begin{eqnarray}
    K_{+}(\xi_{1,1}) + h_4 K_{+}(\xi_{2,1}) &=&0\\
    R_1 K_{-}(\xi_{2,1}) + h_4 R_2 K_{-}(\xi_{2,1})&=&0,
  \end{eqnarray}
  and, therefore, the secular equation
  \begin{eqnarray}
    \frac{K_{+}(\xi_{1,1})}{K_{+}(\xi_{2,1})} =   \frac{R_1}{R_2}  \frac{K_{-}(\xi_{1,1})}{K_{-}(\xi_{2,1})}.
    \label{eq:sec_int_edge}
  \end{eqnarray}
  Again, the coefficients $h_2$ and $h_4$ are then found imposing the same boundary conditions for a fixed energy.

  \section{Meaning of spin in topological insulators\label{app:spin}}
  The bands of a semiconductor, in the presence of spin-orbit coupling,  
  are characterized by their \emph{real} total angular momentum (meaning orbital plus real spin).
  For our topological insulator we can express the relevant bands at the $\Gamma$-point as~\cite{novik2005}
  \begin{eqnarray*}
    \left|\Gamma_6,\pm\frac{1}{2}\right\rangle &\equiv& S \left|\pm \frac{1}{2}\right\rangle \\
    \left|\Gamma_8,\pm\frac{3}{2}\right\rangle &\equiv& \frac{1}{\sqrt{2}}(X\pm i Y) \left|\pm \frac{1}{2}\right\rangle \\  
    \left|\Gamma_8,\pm\frac{1}{2}\right\rangle &\equiv&   
    \pm \frac{1}{\sqrt{6}}\left[ (X\pm i Y) \left|\mp \frac{1}{2}\right\rangle \mp 2Z \left|\pm \frac{1}{2}\right\rangle\right]     \\ 
    \left|\Gamma_7,\pm\frac{1}{2}\right\rangle &\equiv& 
    \frac{1}{\sqrt{3}}\left[ (X\pm i Y) \left|\mp \frac{1}{2}\right\rangle \pm Z \left|\pm \frac{1}{2}\right\rangle\right],  
  \end{eqnarray*}
  where each one of the four pairs obeys 
  $\left|\Gamma_i,-\frac{1}{2}\right\rangle = -\hat{T}\left|\Gamma_i,+\frac{1}{2}\right\rangle$, 
  with $\hat{T}$ the time-reversal operator.   
  The states considered in the effective 4-band model are described as~\cite{konig2008}
  \begin{eqnarray}
    \left|E_1 \pm\right\rangle &=& \alpha \left|\Gamma_6,\pm\frac{1}{2}\right\rangle 
    + \beta  \left|\Gamma_8,\pm\frac{1}{2}\right\rangle\nonumber\\
    \left|H_1 \pm\right\rangle &=& \left|\Gamma_8,\pm\frac{3}{2}\right\rangle
    \label{eq:gamma}
  \end{eqnarray} 
  with $|E_1 \pm\rangle$ and $|H_1 \pm\rangle$ two set of Kramer's partners.
  We can therefore explicitly express the spinor on the block $\tau=1$ as
  \begin{equation}
    |+\rangle = \Psi_1 \left|E_1+\right\rangle + \Psi_2 \left|H_1+\right\rangle
    \label{eq:up}
  \end{equation}
  and similarly we can express its time reversal (we use here the unsymmetrized basis)
  \begin{equation}
    |-\rangle = \hat{T} |+\rangle=  -\Psi_1^*\left|E_1-\right\rangle -\Psi_2^*\left|H_1-\right\rangle.
    \label{eq:down}
  \end{equation}
  We note that the two blocks $\tau=\pm1$, although often referred to as spin up and spin down blocks, 
  do not exactly correspond to eigenstates up and down of the \emph{real} spin along the $Z$ axis.
  We can indeed, using Eq.~(\ref{eq:gamma}), calculate the \emph{real} spin expectation value of these two states, obtaining 
  \begin{eqnarray}
  \left\langle+\right|\hat{\vec{S}}\left|+\right\rangle &=&  
  \frac{\beta}{\sqrt{3}} \left( \Psi_2^* \Psi_1 + \Psi_1^* \Psi_2 \right) \hat{X}+\nonumber\\
  &+& \frac{i \beta}{\sqrt{3}} \left(- \Psi_2^* \Psi_1 + \Psi_1^* \Psi_2 \right) \hat{Y}+\nonumber\\
  &+&  \left(|\Psi_2|^2 \left[1+ |\alpha|^2\right] + \frac{|\Psi_1 \beta|^2}{3}\right) \hat{Z},\label{eq:axes}\nonumber\\
  \langle-|\hat{\vec{S}}|-\rangle &=&  - \langle+|\hat{\vec{S}}|+\rangle.
\label{eq:spin}
\end{eqnarray}
It is now clear that if we choose a spin quantization axis along $\langle+|\hat{\vec{S}}|+\rangle$, 
the orthogonal $|\pm\rangle$ states will correspond to spin up and spin down eigenstates, respectively.
Therefore, in principle, particles injected with spin along this direction will enter only $|+\rangle$ states, 
conversely the ones with opposite spin will enter only $|-\rangle$ states.
However the spin quantization direction depends implicitly on energy through $\Psi_1$ and $\Psi_2$.
In conclusion, in a small enough energy range we can consider $|\pm\rangle$ as spin states 
along the spin quantization axis given by Eq.~(\ref{eq:spin})
and it is meaningful to think of spin injection or detection processes, 
where the TI ring is coupled to external spin-selective materials like ferromagnetic 
contacts or helical edge states of a two dimensional TI.


\bibliography{TopIns}

\begin{thebibliography}{41}
\bibitem[{not()}]{note}
\bibinfo{note}{The real total angular momentum operator
  $\hat{J}_z=\hat{l}_z+\hbar \tau_z -\hat{S}_z$, which commutes with both $H$
  and $H_R$, could have also been used to arrive at the same conclusions.}
\end{thebibliography}
\end{document}